\documentclass[final,5p,times,twocolumn]{elsarticle}

\usepackage{graphicx}
\usepackage{siunitx}
\usepackage{color}

\usepackage{amssymb,amsmath}
\usepackage{amsthm}

\makeatletter
\renewcommand\paragraph{\@startsection{paragraph}{4}{\z@}%
            {-2.5ex\@plus -1ex \@minus -.25ex}%
            {1.25ex \@plus .25ex}%
            {\normalfont\normalsize}}
\makeatother
\newcommand{\midtilde}{\raise.17ex\hbox{$\scriptstyle\mathtt{\sim}$}}
\def \figwidth {0.47}

\journal{Elsevier}

\begin{document}

\begin{frontmatter}



\title{FPGA-based Trigger System for the LUX Dark Matter Experiment}
%%\textcolor{red}{Preliminary draft. ONLY for internal use!}}

%\author[ece,pas]{Eryk Druszkiewicz}
%\ead{eryk.druszkiewicz@rochester.edu}
%
%\author[pas]{Frank Wolfs}
%\ead{wolfs@pas.rochester.edu}
%\address[pas]{Department of Physics and Astronomy, University of Rochester, Rochester, NY }
%\address[ece]{Department of Electrical and Computer Engineering, University of Rochester, Rochester, NY }
\cortext[cor1]{Corresponding author: eryk.druszkiewicz@rochester.edu}
\author[CWRU,SLACNAL,KIPAC]{D.S.~Akerib} %% Joined 4/2008
\author[ICL]{H.M.~Ara\'{u}jo} %% Joined 4/2012, Author on 4/2013
\author[SDSMT]{X.~Bai} %% Joined 7/2009
\author[ICL]{A.J.~Bailey} %% Joined 1/2013, Author on 1/2014
\author[UM]{J.~Balajthy} %% Joined 7/2012, Author on 7/2013
\author[SUPA]{P.~Beltrame} %% Joined 10/2013, Author on 10/2014.
\author[YU]{E.P.~Bernard} %% Joined 1/2010, Author on 1/2011
\author[LLNL]{A.~Bernstein} %% From the start
\author[CWRU,SLACNAL,KIPAC]{T.P.~Biesiadzinski} %% Joined 10/2013, Author on 10/2014
\author[YU]{E.M.~Boulton} %% Joined 3/2013, Author on 3/2014
% ...
%% Joined 2/2009

\address[CWRU]{Case Western Reserve University, Dept. of Physics, 10900 Euclid Ave, Cleveland OH 44106, USA} %% 1
\address[SLACNAL]{SLAC National Accelerator Laboratory, 2575 Sand Hill Road, Menlo Park CA 94205-7015, USA} %% 2
\address[KIPAC]{Kavli Institute for Particle Astrophysics and Cosmology, Stanford University, 452 Lomita Mall, Stanford, CA 94309, USA} %% 3
\address[ICL]{Imperial College London, High Energy Physics, Blackett Laboratory, London SW7 2BZ, UK} %% 4
\address[SDSMT]{South Dakota School of Mines and Technology, 501 East St Joseph St., Rapid City SD 57701, USA} %% 5
\address[UM]{University of Maryland, Dept. of Physics, College Park MD 20742, USA} %% 6
\address[SUPA]{SUPA, School of Physics and Astronomy, University of Edinburgh, Edinburgh, EH9 3FD, UK} %% 7
\address[YU]{Yale University, Dept. of Physics, 217 Prospect St., New Haven CT 06511, USA} %% 8
\address[LLNL]{Lawrence Livermore National Laboratory, 7000 East Ave., Livermore CA 94551, USA} %% 9
% ...
%% 21

\begin{abstract}
LUX is a two-phase (liquid/gas) xenon time projection chamber designed to detect nuclear recoils resulting from interactions with dark matter particles.
Signals from the detector are processed with an FPGA-based digital trigger system that analyzes the incoming data in real-time, with just a few microsecond latency.
The system enables first pass selection of events of interest based on their pulse shape characteristics and 3D localization of the interactions. 
It has been shown to be $>$99\% efficient in triggering on S2 signals induced by only few extracted liquid electrons. It is continuously and reliably operating since its full underground deployment in early 2013. This document is an overview of the systems capabilities, its inner workings, and its performance.

\end{abstract}


\end{frontmatter}

%%\linenumbers
%%\linenumbers
\section{Introduction}
\begin{figure}[!ht]
\centering
		\includegraphics[width=0.48\textwidth]{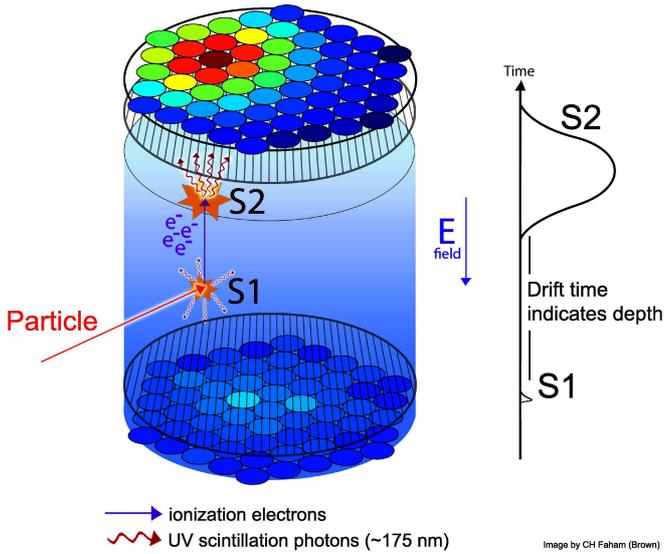}
\caption[Detector]{Depiction of signal generation in the LUX detector~\cite{JChapmanDAQ}. Upon an interaction with the xenon medium, a prompt primary light signal (S1) is followed by a wider secondary light signal (S2), with appropriate time separation due to the drift length of electrons to the gas phase of the detector.}
\label{fig:Detector}
\end{figure}

The Large Underground Xenon (LUX) detector is located at the Sanford Underground Research Facility (SURF) at 4850~ft underground. LUX is designed to detect potential dark matter candidates called Weakly Interacting Massive Particles (WIMPs)~\cite{LUXNimPaper}. It utilizes a two-phase (liquid/gas) time projection chamber (TPC) filled with 350 kg of cryogenically cooled xenon~\cite{XenonReviewAprileDoke,XenonReviewChepelAraujo}. As shown in Fig.~\ref{fig:Detector}, the incoming particles that interact with xenon atoms induce primary scintillation light,  called S1 light, and ionization electrons. From the point of interaction, the ionization electrons drift towards the surface in a uniform electric field, generated by gate (top) and cathode (bottom) wire grids. Once the electrons reach the liquid surface, they are extracted into the gas region by a strong electric field. As the electrons are accelerated in the xenon gas, they create electroluminescence, called S2 light. The relation between the S1 and S2 light enables identification of the type of interaction that generated them.
\begin{figure*}[!ht]
\centering
		\includegraphics[width=0.8\textwidth]{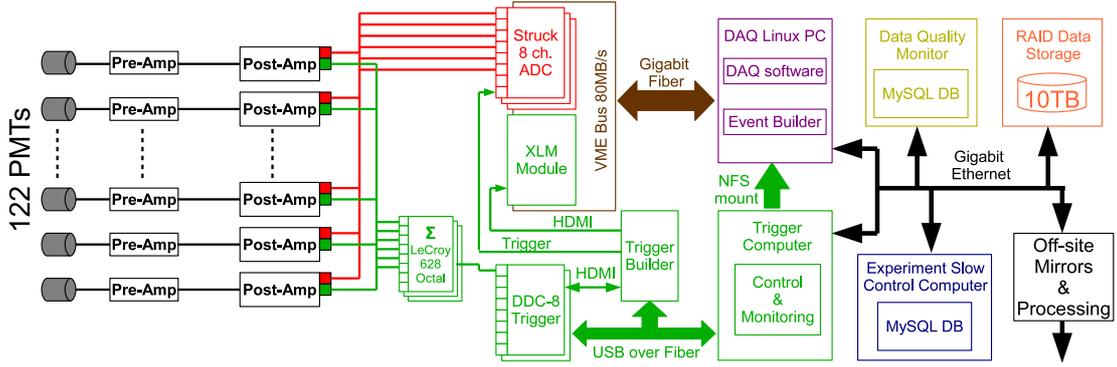}
\caption[DAQSystem]{Overview of the signal and data flow in LUX. The PMT signals are pre-amplified immediately after leaving the xenon space. Post-amplifiers give the signals their final gain and anti-aliased shape, before they are fed into the Struck and Trigger digitizers~\cite{ JChapmanDAQ,citeStruck}. The trigger informs the Strucks about events of interest. Reduced quantities of the trigger system are merged with the waveform data stream (via XLM module) and stored for off-line verification and analysis.}
\label{fig:DAQSystem}
\end{figure*}
\\\indent The signals are detected and amplified with 122 photomultiplier tubes (PMTs), split equally between arrays at the top and bottom of the TPC~\cite{JChapmanDAQ}. Figure~\ref{fig:DAQSystem} shows an overview of the LUX signal and data flow. The PMT signals are shaped and amplified by Pre- and Post-Amplifiers. The individual PMT channels are summed into 16 groups before reaching the Trigger System. PMTs are set to generate about \num{3.3e6} electrons per photoelectron (phe), which at the PMT output corresponds to a pulse with an area of \midtilde13.3~mVns. The analog chain amplification and shaping increases single phe pulse area to \midtilde151~mVns and FWTM of 54~ns at the trigger digitizers.
\\\indent The first WIMP search based on the analysis of 85.3~live-days of data with a fiducial volume of 118~kg, allowed for setting a limit on spin-independent WIMP-nucleon elastic scattering with a minimum upper limit on the cross section of $7.6 \times 10^{-46}$~cm$^{2}$ at a WIMP mass of 33~GeV/c$^2$~\cite{LUXFirstResults}. Currently the LUX experiment is in its 300-day run, that is due to conclude mid 2016 and will further improve the limit.
\\\indent This paper presents an overview of the LUX trigger system, which is required to make real-time event selection based on the S1 and S2-type signals. It is required to be sensitive to single phe and single liquid electron signals. To keep the dead time of the system, due to estimated PMT dark rate, at a 1\% level or better, the selection decisions are required to be made within 10~$\mu$s window. The system was designed with more features than are being used in the actual WIMP-search run (Section~\ref{subsec:Run03WIMPSearch}). 

%\\\indent This paper presents an overview of the LUX trigger system that allows for real-time event selection based on the S1 and S2-type signals. The system was designed with more features than are being used in the actual WIMP-search run (Section~\ref{subsec:Run03WIMPSearch}).

\section{Trigger System Overview}

\begin{figure}[!ht]
\centering
		\includegraphics[width=\figwidth\textwidth]{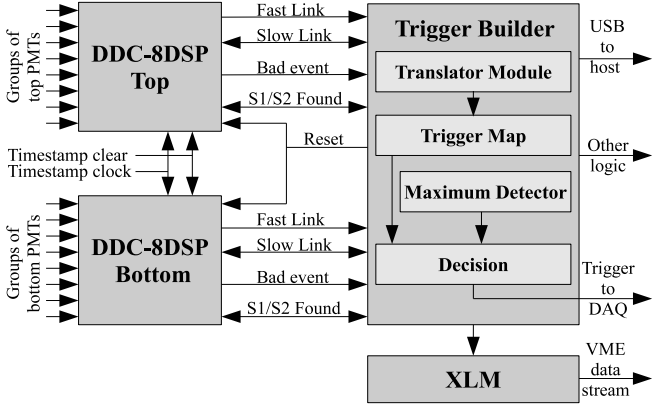}
\caption[BoardComm]{Schematic diagram of the LUX trigger. The TB interacts closely with the DDC-8DSPs as well as other DAQ components such as the XLM module. Trigger decisions and reduced quantities are timestamped using external 100~MHz DAQ clock. Details are provided in the main text.}
\label{fig:BoardComm}
\end{figure}

\begin{figure*}[!ht]
\centering
		\includegraphics[width=0.84\textwidth]{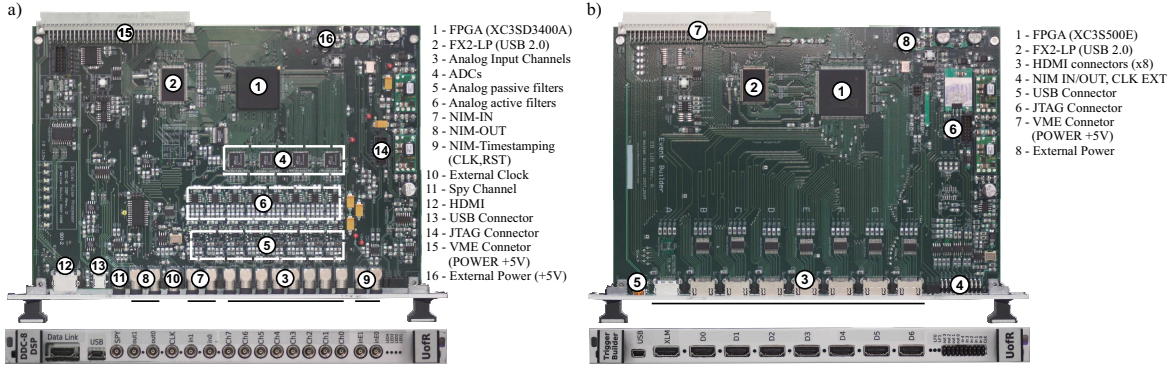}
\caption[DDC8andTriggerBuilder]{a) The DDC-8DSP captures data with 14-bit resolution at 64MHz and processes it with a Xilinx Spartan-3A DSP FPGA. b) The Trigger Builder uses a Spartan-3E 500 Xilinx FPGA and enables second level processing of data received over high speed HDMI links. It allows making sophisticated trigger decisions based on information from both top and bottom PMTs.}
\label{fig:DDC8andTriggerBuilder}
\end{figure*}

The LUX trigger is a powerful system that has the ability to use the pulse shape information of the PMT signals to identify potential dark matter events and reject background events (e.g.~events outside an inner fiducial volume, events with large energy depositions). Having the ability to differentiate between S1 and S2 signals, the trigger system can accept events in modes that use just S1 signal information, just S2 signal information, or the combination of the two. Pattern recognition is used to select events of interest based on the time, energy, and spatial information provided by the PMTs. Background events associated with large S1 and/or S2 signals and/or invalid geometrical patterns can be vetoed. 
The system is designed to make the trigger decision within as little as 1~$\mu$s from the time an interaction of interest occurs. Real-time processing is made possible thanks to FPGA technology. When used correctly, it is dependable, flexible, and cost effective, especially for projects requiring iterative development. 
%The system is designed to make the trigger decision within as little as 1~$\mu$s from the time an event of interest arrives. Real-time processing is made possible thanks to FPGA technology. When used correctly, it is dependable, flexible, and cost effective, especially for projects requiring iterative development. 
\\\indent The analog sums of seven to eight PMTs are sent to the trigger system, as shown in Fig.~\ref{fig:BoardComm}. The summed signals are digitized and processed by two DDC-8DSP modules (Fig.~\ref{fig:DDC8andTriggerBuilder}a)~\cite{Skutek}. The digitization is done at 64 MHz with 14-bit resolution. The two DDC-8DSP modules communicate with the Trigger Builder (TB) (Fig.~\ref{fig:DDC8andTriggerBuilder}b). The TB is used to make more advanced trigger decisions using information from both top and bottom PMTs. The Fast and Slow Links use HDMI cables and the rest of the signals travel over regular LEMO cables. The fast unidirectional link uses four differential pairs (LVDS) of the HDMI standard and is used to transfer data blocks, such as S1 and S2 hit vectors, waveforms, timestamps, and other reduced quantities. The slow link uses seven single-ended, bidirectional connections of the HDMI standard and is used to communicate the finite state machine states across the boards.

%\begin{figure}[!ht]
%\centering
		%\includegraphics[width=\figwidth\textwidth]{DDC8DSP.eps}
%\caption[DDC8DSP]{The DDC-8DSP captures data with 14-bit resolution at 64MHz and processes it with a Xilinx Spartan-3A DSP FPGA.}
%\label{fig:DDC8DSP}
%\end{figure}

%\begin{figure}[!ht]
%\centering
		%\includegraphics[width=\figwidth\textwidth]{TriggerBuilder.eps}
%\caption[TriggerBuilder]{The Trigger Builder uses a Spartan-3E 500 Xilinx FPGA and enables second level processing of data received over high speed HDMI links. It allows making sophisticated trigger decisions based on information from both top and bottom PMTs.}
%\label{fig:TriggerBuilder}
%\end{figure}

\indent The TB has one HDMI connection dedicated to send operational, configuration, and diagnostic information to an Extended Logic Module (XLM)~\cite{JTec}, located in the VME crate of the DAQ system. This allows the trigger information to be merged with the DAQ data stream for off-line cross-checks and analysis. Since the DAQ and Trigger systems run off 100~MHz and 64~MHz clocks, respectively, the trigger decisions and reduced quantities are time-stamped using the DAQ clock, using dedicated high-speed DSP48A slices on the FPGA ~\cite{XilinxDSP48A}.
\\\indent The DDC-8DSPs and the TB are both controlled via a USB 2.0 interface using a dedicated host computer. The USB 2.0 communication is deployed using USB-over-Fiber~\cite{Adnaco} for performance and minimizing the potential for ground loops.

\section{Filters}
\subsection{Analog filters}

\subsubsection{Butterworth and $2^{nd}$-order Bessel Filters}
The DDC-8DSP has two stage filtering at each analog input. A passive Butterworth filter provides initial signal conditioning. The Butterworth filter is followed by a $2^{nd}$-order Bessel filter. This is an active filter which was designed as an anti-aliasing filter that has minimal overshoot characteristics. Both filters have been tuned to have a corner frequency of 24~MHz.

\subsubsection{Spice simulations}
\begin{figure}[!ht]
\centering
		\includegraphics[width=\figwidth\textwidth]{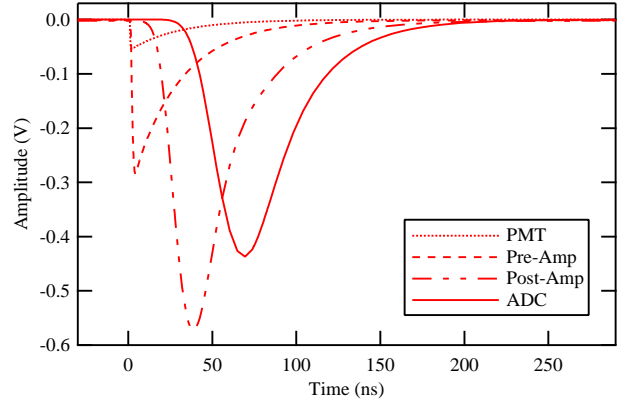}
\caption[SPiceOut]{A 200 photoelectron ideal S1 pulse propagated through the LTSpice model. The raw PMT signal (PMT), Pre-Amplifier output (Pre-Amp), Post-Amplifier DDC-8DSP output (Post-Amp), and the DDC-8DSP ADC input (ADC) are shown.}
\label{fig:SPiceOut}
\end{figure}

In order to predict the trigger performance, the entire electronics chain was modeled using LTSpice~\cite{LTSpice}. 
%The top overview of the model is shown in Fig.~\ref{fig:SPICEModel}. 
Figure~\ref{fig:SPiceOut} shows an example of the output of the LTSpice simulations.
%
%\begin{figure}[!ht]
%\centering
		%\includegraphics[width=\figwidth\textwidth]{SPICEModel.eps}
%\caption[SPICEModel]{Top-level full chain SPICE model. X1, X2 are internal to the DDC-8DSP and represent the Butterworth input filter with baseline level adjustment and the $2^{nd}$-order Bessel filter, respectively.}
%\label{fig:SPICEModel}
%\end{figure}

\subsection{Digital filters}
\begin{figure}[!ht]
	\centering
		\includegraphics[width=0.48\textwidth]{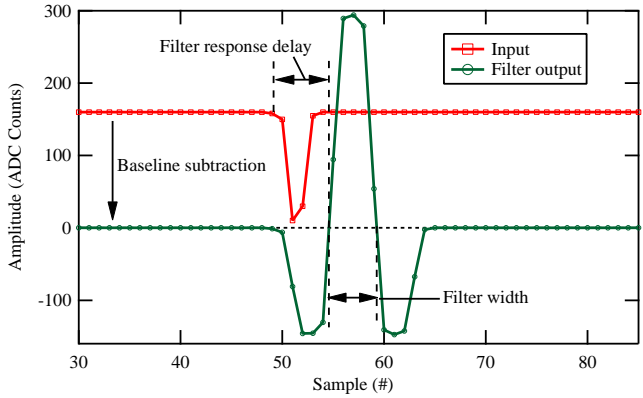}
	\caption[Digital Filter Response]{Sample response of S1 filter with width (\textit{n}) set to 4. The filter performs baseline subtraction and its response is proportional to the area of the input pulse.}
	\label{fig:FilterResponse}
\end{figure}

The digitized input signals are processed on the FPGA. The digital filter implemented on the FPGA is an FIR filter which is defined by the following general equation:

\begin{equation}
h(i)=B\left(\sum^{i-(m+n)}_{j=i-(2n+m-1)}x(j)\right)-A\left(\sum^{i-n}_{j=i-(m+n-1)}x(j)\right)+B\left(\sum^{i}_{j=i-(n-1)}x(j)\right).
\label{eq:ising}
\end{equation}
The parameters \textit{m} and \textit{n} define the sample width of the main and side filter lobes, respectively. The \textit{A} and \textit{B} parameters are the filter  coefficients for the main and side lobes, respectively. In order for the filter to eliminate baseline variations, the filter coefficients must have a total weight of zero. This requires that $A~\cdot~m~=~2~\cdot~B~\cdot~n$. Figure~\ref{fig:FilterResponse} shows the filter response to an input pulse. The filter acts as an integrator and its response is proportional to the area of the input pulse when \textit{m} is tuned to the signal width.

\subsubsection{S1 Filter}
\begin{figure}[!ht]
\centering
		\includegraphics[width=0.25\textwidth]{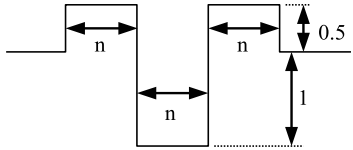}
\caption[S1FilterShape]{Depiction of the S1 filter with a main to side lobe ratio of 1:1.}
\label{fig:S1FilterShape}
\end{figure}

The S1 filter is adjusted to detect S1-like pulses. It is defined by Eq.~\eqref{eq:ising} with parameters $A = 1$, $B = 0.5$, and $m = n$, and it is shown in Fig.~\ref{fig:S1FilterShape}. The filter width (\textit{n}) is programmable and can be 1 to 16 samples, which on a 64-MHz platform corresponds to a filter width range of 15.625 to 250~ns. This range is sufficient for the trigger to encapsulate a typical S1-like pulse which has a FWHM of \midtilde80 ns.
Although the filter reduces the effect of baseline variations, it increases the noise associated with summing multiple digitized samples. The noise observed at the ADC output is increased by a factor expressed by Eq.~\eqref{eq:FGain} 
\begin{equation}
%%\sqrt{(\frac{1}{2})^{2}n + (1)^{2}m + (\frac{1}{2})^{2}n} = \sqrt{\frac{3}{2}n}.
\sqrt{{A}^{2}n + {B}^{2}m + {A}^{2}n} = \sqrt{2{A}^{2}n + {B}^{2}m}
\label{eq:FGain}
\end{equation}
%%The reduction in baseline variation noise far surpasses the raise due to the random fluctuations.
and for the S1 filter equals $\sqrt{\frac{3}{2}n}$.

\begin{figure}[!ht]
\centering
		\includegraphics[width=0.48\textwidth]{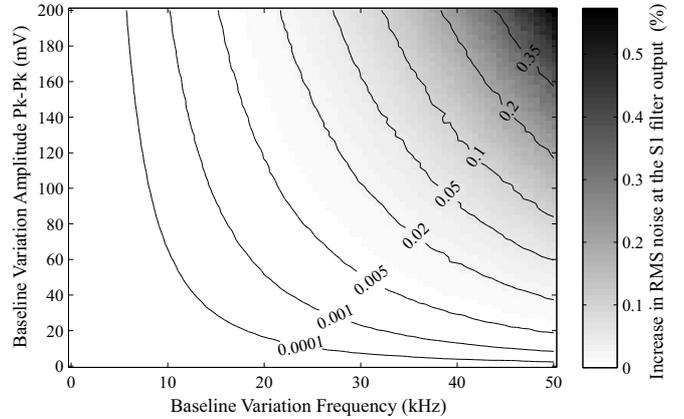}
\caption[S1NoiseImmunityv1]{Immunity of S1 filter (n~=~4) to sinusoidal baseline variations.}
\label{fig:S1NoiseImmunityv1}
\end{figure}

\indent The filter efficiency to reject relatively slow baseline variations was estimated in the following way. We recorded noise waveforms of the analog chain as seen by the ADCs on the DDC-8DSP boards, passed them through the S1 filter (n~=~4), and measured the RMS noise at its output. Then we took the same noise waveforms, added sinusoidal signals with known amplitude and frequency, passed them through the S1 filter, and measured the RMS noise at the filter output. The resulting change in RMS noise as a function of the amplitude and frequency of the sinusoidal baseline component is shown in Fig.~\ref{fig:S1NoiseImmunityv1}.
The increase in noise at the filter output is very small over a broad range of amplitudes and frequencies. It reaches just above 0.5\% for relatively big superimposed baseline variation of 200~mV (pk-pk) and 50~kHz. 

\paragraph{S1-like pulse generation}

\indent In order to evaluate the performance of the S1 filter, one needs to be able to simulate S1-like pulses at the PMT output, which can be passed into the Spice model and analyzed further. Analyzing actual PMT output signals ~\cite{PSorensen,LDeViveiros} we conclude that single phe pulses can be safely approximated with a Gaussian function, but as the number of phes grows within the pulse, the overall shape morphs to one expressed by the following envelope:  
\begin{equation}
S1_{Envelope} =A\cdot \left(\exp\left(\frac{-t+t_{0}}{\tau_{1}}\right)-\exp\left(\frac{-t+t_{0}}{\tau_{2}}\right)\right)\cdot u_{0}(t-t_{0}).
\label{eq:S1Envelope}
\end{equation}
Measurements showed that S1 pulses are bounded by a pulse shape with a \midtilde5.5~ns rise-time and a \midtilde29~ns fall-time. The rise-time is defined as the time required to go from 10\% to 90\% of the pulse amplitude. The fall time is defined as the time from the peak of the pulse to 1/$e$ of the peak value, which closely matches the triplet decay time in liquid xenon~\cite{PSorensen}. The values of $\tau_{1}$ and $\tau_{2}$ in Eq.~\ref{eq:S1Envelope}, adjusted to reproduce the measured rise and fall times are, \midtilde5~ns and \midtilde23~ns, respectively.
\begin{figure}[!ht]
\centering
		\includegraphics[width=\figwidth\textwidth]{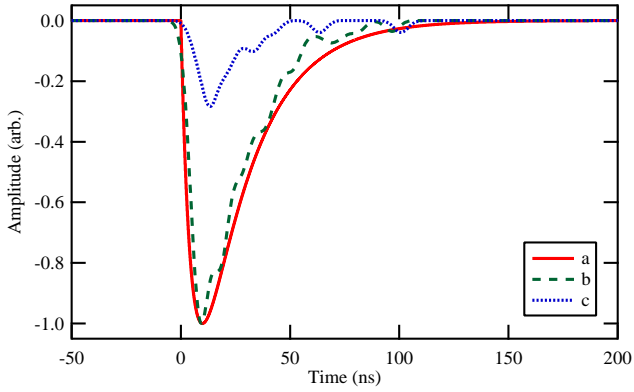}
\caption[S1PulseShape]{a - S1 pulse shape envelope with a 5.5 ns rise-time and a 29 ns fall-time. b- Sample S1 induced by 100 photoelectrons. c - Sample S1 induced by 20 photoelectrons.}
\label{fig:S1PulseShape}
\end{figure}
In order to generate a multi-phe S1-like pulse with an envelope as shown in Fig.~\ref{fig:S1PulseShape} one needs to superimpose a number of single phes with appropriate time offsets, such that, as the number of phes grows, the resulting pulse envelope has a shape similar to that observed in the experimental data. To do that one needs to take the pulse envelope, normalize its area to unity, and perform a cumulative sum on it. The (0,1) range of the cumulative sum yields a probability mapping that allows us to use a uniform random number generator to obtain a set of offsets for the construction of a large S1 like pulse.

%Plotting the result versus respective (0,1) range (Fig.~\ref{fig:UniformToOffsetCDF}) yields a probability mapping that allows us to use a uniform random number generator to obtain a set of offsets for construction of a large S1 like pulse.
%\begin{figure}[!ht]
%\centering
		%\includegraphics[width=\figwidth\textwidth]{UniformToOffsetCDF.eps}
%\caption[UniformToOffsetCDF]{Mapping between a uniform distribution and positioning of the individual single phes within a multi-phe S1-like pulse}
%\label{fig:UniformToOffsetCDF}
%\end{figure} 

\begin{figure}[!ht]
\centering
		\includegraphics[width=\figwidth\textwidth]{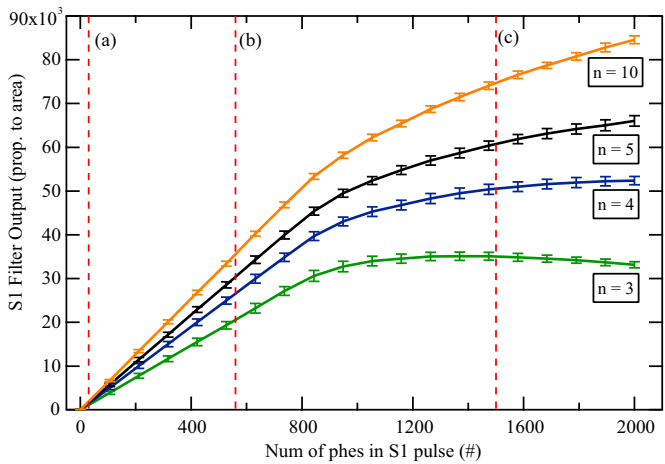}
\caption[S1DynamicRange]{S1 filter dynamic range for a single DDC-8DSP channel for different filter widths (\textit{n}). For reference: a) 30 phe S1 upper cut for Run3 WIMP search~\cite{LUXFirstResults}, b) The full $^{137}\textrm{Cs}$ energy deposition peak from the 662 keV $\gamma$-rays appears at 5624 phe (562 phe per trigger channel of the bottom PMTs)~\cite{LUXTechnicalResultsFromSurface}, c)~S1 pulses from $\gamma$ ray events reach up to \num{1.5e4} phe (1500 phe per trigger channel of the bottom PMTs)~\cite{LUXBackgroundPaper}. }
\label{fig:S1DynamicRange}
\end{figure}

\paragraph{Dynamic Range}
S1-like pulses with a given number of phes are passed through the model of the analog chain and the S1 filter. By recording the maximum output values of the S1 filter as a function of the filter width \textit{n}, a dynamic range can be determined, as is shown in Fig.~\ref{fig:S1DynamicRange}.
The results of the simulations for a nominal PMT gain of \num{3.3e6} show that we can expect linearity up to about 800 phes. This range is relatively independent of the filter length. Because of the monotonicity of the filter response for $n \geq 4$, one can still perform pulse area discrimination in the non-linear region. For shorter filters (e.g. \textit{n} = 3) there is a slight falloff for the S1 filter output for pulses above 1400 phes. This is related to short filters being more sensitive to the side-effects of progressively larger saturation of analog input stages. 

\paragraph{Filter sensitivity}
\label{sec:Filtersensitivity}
In order to estimate the lower threshold that can be applied to the S1 filter output we carried simulations that evaluate the following:
\begin{itemize}
	\item False trigger rate -- determine how many times per second the S1 filter will identify a noise fluctuation in the baseline as a valid pulse for a given threshold.
	\item Trigger efficiency -- estimate what fraction of single phe pulses will cross the lower threshold of the S1 filter and thus be identified as a valid S1-like pulse.
\end{itemize}
In all our simulations we have used real noise baseline sets collected at SURF. The sets represent the electronic noise induced on one trigger channel when eight PMT bases were fully biased and the signal was passed through the entire analog chain to the trigger input. A few thousand waveforms, each containing 128~K samples, were captured with the DDC-8DSPs, totaling four seconds worth of noise waveforms. 
The simulations take into account the PMT single phe response which can be reasonably approximated by a Gaussian distribution with a sigma of 37\%.

\begin{figure}[!ht]
\centering
		\includegraphics[width=\figwidth\textwidth]{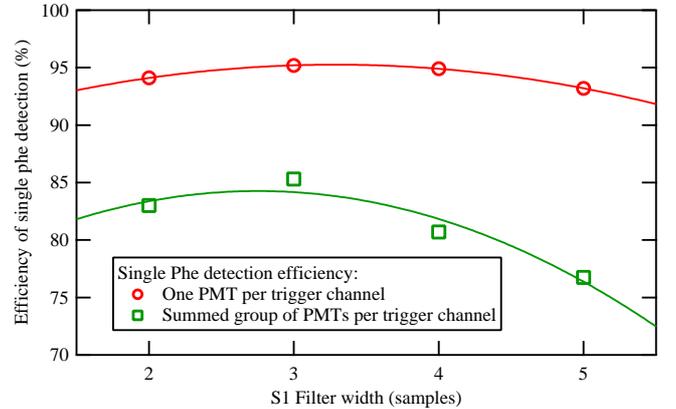}
\label{fig:SpheEfficiencyVsS1L}
\caption[SpheEfficiencyVsS1L]{Efficiency of the trigger S1 filter in detecting single photoelectrons when accepting 1 Hz of false triggers. Having to sum multiple PMT signals for the trigger channels decreases the detection efficiency by \midtilde15\%.}
\end{figure}
Simulation results shown in Fig.~\ref{fig:SpheEfficiencyVsS1L} indicate that an S1 filter width (\textit{n}) of three samples yields the best single phe detection efficiency, but at the price of the dynamic range as shown earlier in Fig.~\ref{fig:S1DynamicRange}. Finally, a width of four samples has been determined to be optimal. If individual PMTs were connected to individual trigger channels, we could expect a up to 95\% efficiency in detecting single photoelectrons with a \midtilde1~Hz of false triggers. However, summing PMT channels increases the level of the baseline noise and forces us to raise the threshold to preserve the \midtilde1~Hz of false triggers. According to the simulations the higher threshold causes a drop in efficiency of detecting single phes to \midtilde80\%. 

\subsubsection{S2 Filter}
\begin{figure}[!ht]
\centering
		\includegraphics[width=0.30\textwidth]{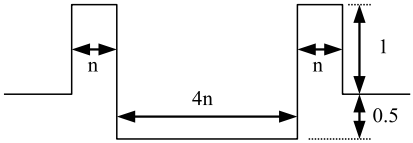}
\caption[S2FilterShape]{S2 filter with a main to side lobe ratio of 4:1, which saves 50\% of FPGA pipeline resources, and enables better separation of adjacent pulses.}
\label{fig:S2FilterShape}
\end{figure}

The S2 signals are wider than the S1 signals and thus require an appropriately longer filter. The filter shape is different from the S1 filter. It is still defined by Eq.~\eqref{eq:ising}, but with $A = 1$, $B = 2$ and $m = 4 \cdot n$. The parameter \textit{n} is programmable and can be 1 to 64 samples, which on a 64-MHz platform corresponds to filter width (\textit{m}) range of 62.5~ns to 4~$\mu$s.
%to one described by the following Equation:
%\begin{equation}
%h(i)=\sum^{-5N}_{i=-(6N-1)}x(i)-\frac{1}{2}\sum^{-N}_{i=-(5N-1)}x(i)+\sum^{0}_{i=-(N-1)}x(i)
%\label{eq:S2FilterEquation}
%\end{equation}
For this filter, the ratio of the width of the main lobe to the side lobe is changed from 1:1 to 4:1, as shown in Fig.~\ref{fig:S2FilterShape}. This enables us to save 50\% pipeline delay elements associated with the filters in the FPGA. The four times smaller width of the side lobe enables better discrimination of S2-like pulses which arrive close together, in for example multiple scatter events. If the filter width (\textit{m}) is set to 2~$\mu$s, then with a 1:1 S2 filter the minimal recommended pulse separation (defined as time-at-baseline) is \midtilde2~$\mu$s, while for the 4:1 S2 filter the recommended pulse separation drops down to \midtilde500~ns. These recommended separations come directly from the size of the side lobes (\textit{n}).

\paragraph{S2-like pulse generation}

\begin{figure}[!ht]
\centering
		\includegraphics[width=\figwidth\textwidth]{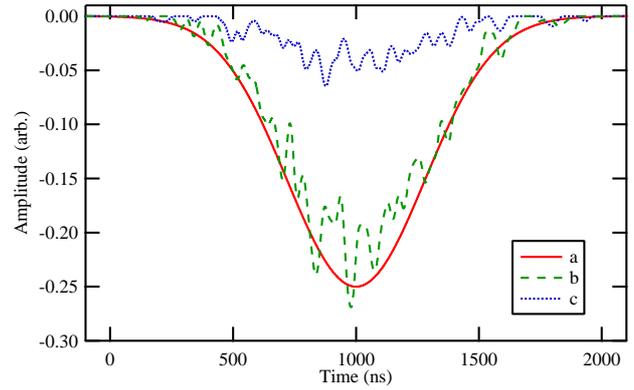}
\caption[S2matlab]{a - S2 pulse envelope with a width of 280~ns (1$\sigma$). b- Sample S2 induced by 1000 photoelectrons. c - Sample S2 pulse induced by 200 photoelectrons.}
\label{fig:S2matlab}
\end{figure}

The signals at the PMT output that are associated with S2-like pulses have been found to be well approximated by the superposition of individual photoelectrons with a Gaussian distribution envelope. The width of the distribution is mostly dependent on the depth of the interaction and the drift field, and a typical value of 280~ns (1$\sigma$) has been chosen for the study. The individual photoelectrons are approximated with a Gaussian shape with a FWHM of 7.7~ns and an area of 13.3~mVns for a PMT gain of \num{3.3e6}. Such S2-like pulses are fed into the simulation model which yields the expected output of the ADCs on the DDC-8DSP modules. A sample of a simulated S2 signal is shown in Fig.~\ref{fig:S2matlab}.

\begin{figure}[!ht]
\centering
		\includegraphics[width=\figwidth\textwidth]{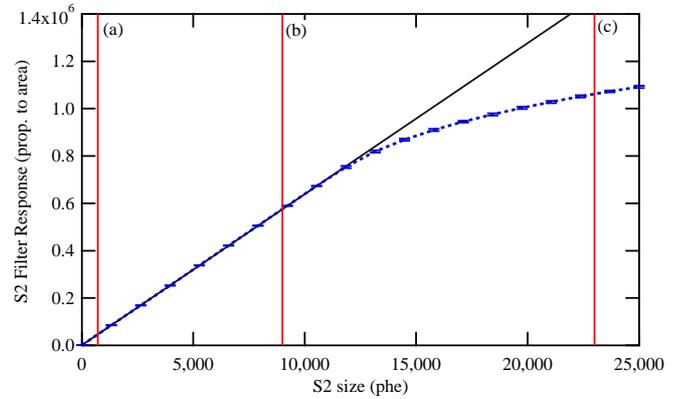}
\caption[S2DynamicRangeonC12]{S2 filter dynamic range for a single DDC-8DSP channel. For reference: a - For Run3 WIMP search the upper limit of S2 total size was 3300 phe (at most \midtilde730 phe in one trigger channel)~\cite{LUXFirstResults}, b - The LUX PMT has a non-linearity of 2\% at 9,000~phe, c - The coupling capacitors of the LUX PMT base fully deplete for pulse size of 23,000~phe~\cite{CarlosThesis}.}
\label{fig:S2DynamicRangeonC12}
\end{figure}

\paragraph{S2 Dynamic Range}
By passing the simulated pulse with a known amount of photoelectrons through the S2 filter and recording the maximum output value, a dynamic range plot can be generated, shown in Fig.~\ref{fig:S2DynamicRangeonC12}. For a single trigger channel, the S2-filter response is linear up to 11,000 photoelectrons. Beyond that, the response is not linear, but pulse area discrimination is still possible.

\begin{figure}[!ht]
\centering
		\includegraphics[width=\figwidth\textwidth]{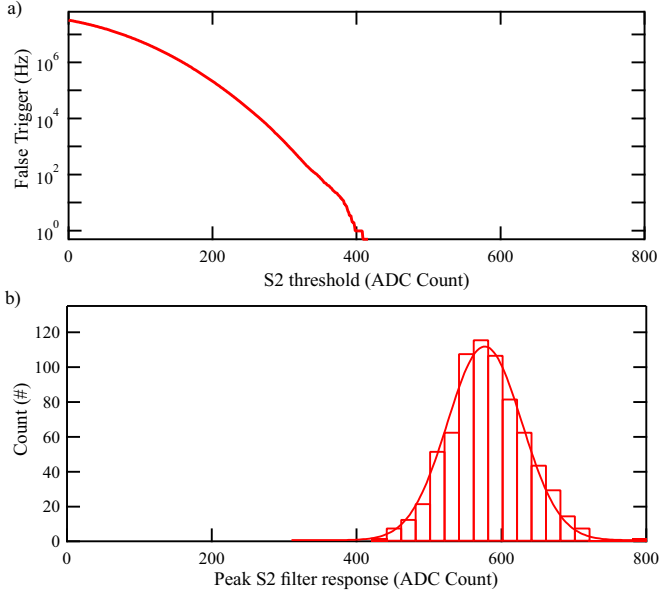}
\caption[S2FalseTriggerand10phe]{a - False Trigger rate of a 2~$\mu$s wide S2 filter in response to a real noise waveform representing a single trigger channel, b - distribution of the peak responses of the S2 filter to six photoelectron S2 input signals seen by a single trigger channel.}
\label{fig:S2FalseTriggerand6phe}
\end{figure}

In order to estimate the lower threshold that can be set on the S2 filter output, the following two elements have been analyzed:
\begin{itemize}
	\item False trigger rate -- estimation of the rate of S2 filter lower threshold crossings due to fluctuations in the baseline.
	\item S2 filter response to very small signals induced by as little as six photoelectrons in a single summed trigger channel.
\end{itemize}
Figure~\ref{fig:S2FalseTriggerand6phe} shows that, with the measured noise levels of summed channels, the S2 filter can detect 99.9\% of six photoelectrons pulses seen by a single trigger channel, with at most 1 Hz of false triggers. A single liquid electron, which is an electron extracted from the surface of the liquid xenon, is expected to produce a total of 20-30 photoelectrons.

\subsubsection{S1 and S2 pulse identification}
\begin{figure}[!ht]
\centering
		\includegraphics[width=0.48\textwidth]{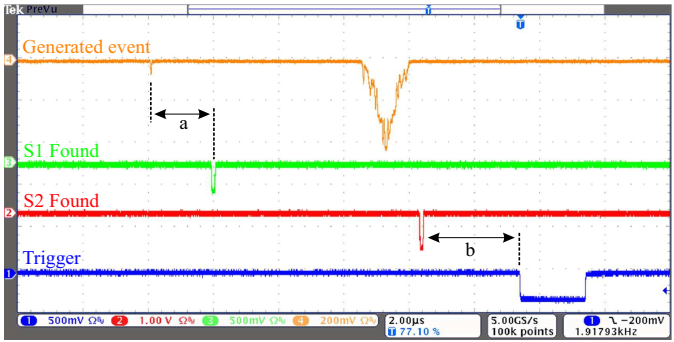}
\caption[SampleScopeEvent]{Generated S1\&S2 event and the trigger response. Ch4 - event waveform. Ch3 - signal indicating an S1 was found, in this case \textit{S1 = S1 not S2}, hence the detection delay. Ch2 - signal indicating an S2 was found. Ch1 - final trigger decision at the Trigger Builder level. Labels \textit{a} and \textit{b} are referenced in section~\ref{sec:RespToSynthEvents}.}
\label{fig:SampleScopeEvent}
\end{figure}

By design, the S1 filter is sensitive to S1-like features of real S2 pulses. To reduce the effect of S1 filter triggering on real S2-like pulses, we have a user selectable option where an \textsl{S1 Found = (S1 filter did and S2 filter did not cross threshold)}. If this option is selected, the S1 filter response is delayed to align it with the S2 filter response and then a requirement of crossing the lower threshold of the S1 filter and not crossing the lower threshold of the S2 filter is imposed. A sample of proper S1 and S2-like pulse identification is shown in Fig.~\ref{fig:SampleScopeEvent}. It should be noted that this approach for S1 pulses whose area exceeds the lower threshold of the S2 filter, will not generate a trigger.

%\begin{figure}[!ht]
%\centering
		%\includegraphics[width=\figwidth\textwidth]{SampleWaveformFilterResponse.eps}
%\caption[FilterResp]{a - Sample real event from the PMTs, b - raw S1 filter response, c - raw S2 filter response.}
%\label{fig:FilterResp}
%\end{figure}

\section{Trigger modes}
The trigger has the capability to differentiate between S1 and S2 signals, and identify events based solely on S1 or S2 signal information, or combined S1 and S2 signal information. The following three trigger modes have been defined: S1Mode, S2Mode and S1\&S2Mode.
\subsection{S1 Mode, S2 Mode}
\begin{figure}[!ht]
\centering
		\includegraphics[width=\figwidth\textwidth]{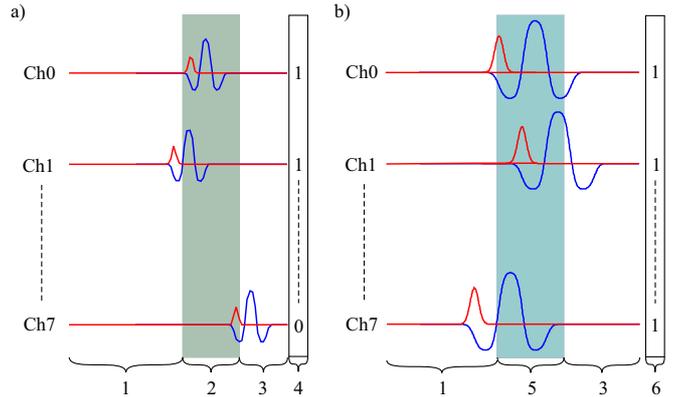}
\caption[S1S2mode]{a) S1Mode, b) S2Mode, 1- Programmable quiet time, 2- S1 coincidence window, 3- programmable hold-off, 4- S1 hit-vector, 5- S2 coincidence window, 6- S2 hit-vector. The red and purple traces depict input signals and their filter responses, respectively.}
\label{fig:S1S2mode}
\end{figure}

S1Mode and S2Mode are similar. Their Finite State Machine (FSM) flow is shown in Fig.~\ref{fig:S1S2mode} except that they search for different pulse types. When the search for an event of interest begins, there is the option to require that the detector is quiet for a certain period of time. During this \textit{quiet time}, no trigger channel can see a signal that crosses the lower threshold of the S1 filters, independent of the trigger mode. The length of the quiet time is user selectable and can range from 0 to 65~ms in 1~$\mu$s steps. The first signal that is detected after the quiet time requirement is satisfied initiates the beginning of the coincidence window. The length of the coincidence window is user selectable and can range from 120~ns to 8~$\mu$s in 32~ns steps. It is separate for S1 and S2. At the end of the coincidence window we record which channels saw their respective filter lower thresholds crossed. We store that information in hit-vectors and pass it to the TB for further processing. We also keep track of which trigger channel generates the maximum S2 filter response which enables more precise fiducialization of events (Section~\ref{subsec:PMTSummAndMaxTrigger}). If a trigger condition is met and a trigger signal is issued to the DAQ system, a hold-off counter is armed with a user-selectable value and starts counting down until reaching zero. While the hold-off counter is counting down, no new triggers are sent to the DAQ. This in many instances prevents triggering on tails of previously triggered interactions. The hold-off time can range from 4~$\mu$s to 65~ms in 1~$\mu$s steps.
\subsection{S1\&S2 Mode}
\begin{figure}[!ht]
\centering
		\includegraphics[width=\figwidth\textwidth]{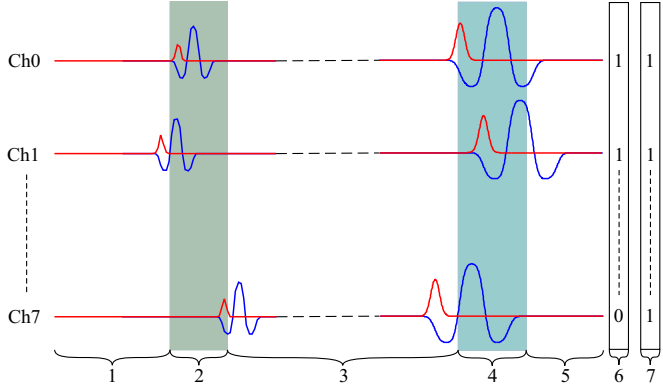}
\caption[S1andS2mode]{S1\&S2 combined mode, 1- quiet time, 2- S1 coincidence window, 3- drift time, 4- S2 coincidence window, 5- hold-off, 6,7- S1 and S2 hit-vectors.}
\label{fig:S1andS2mode}
\end{figure}

S1\&S2Mode is a combined trigger mode where the FSM utilizes information about the two types of pulses simultaneously. This mode glues the two previous separate modes together, by taking into account pulse separation (\textit{drift time}) which allows cuts to be made on vertical position. This is illustrated in Fig.~\ref{fig:S1andS2mode}. The user selectable parameters are the same as described earlier with the addition of a new parameter \textit{drift time} which can range from 15.625~ns to 1.024~ms in 15.625~ns steps. After the S1 coincidence window runs out, the \textit{drift time} counter starts incrementing. If it reaches the user selected value and if no S2-like pulse is found, the trigger issues a reset and starts looking for another event of interest. If an S2-like signal is found before the drift time counter runs out, then the current trigger cycle is carried out to the end (as it would in S2 Mode). At the end of S1\&S2 Mode cycle at the DDC-8DSP level we are left with appropriate S1 and S2 hit-vectors and the record of which channel generated the maximum S2 filter response and what its value was during the cycle. This information is sent to the TB for further processing. 

\section{Trigger Maps}
\begin{figure}[!ht]
\centering
		\includegraphics[width=\figwidth\textwidth]{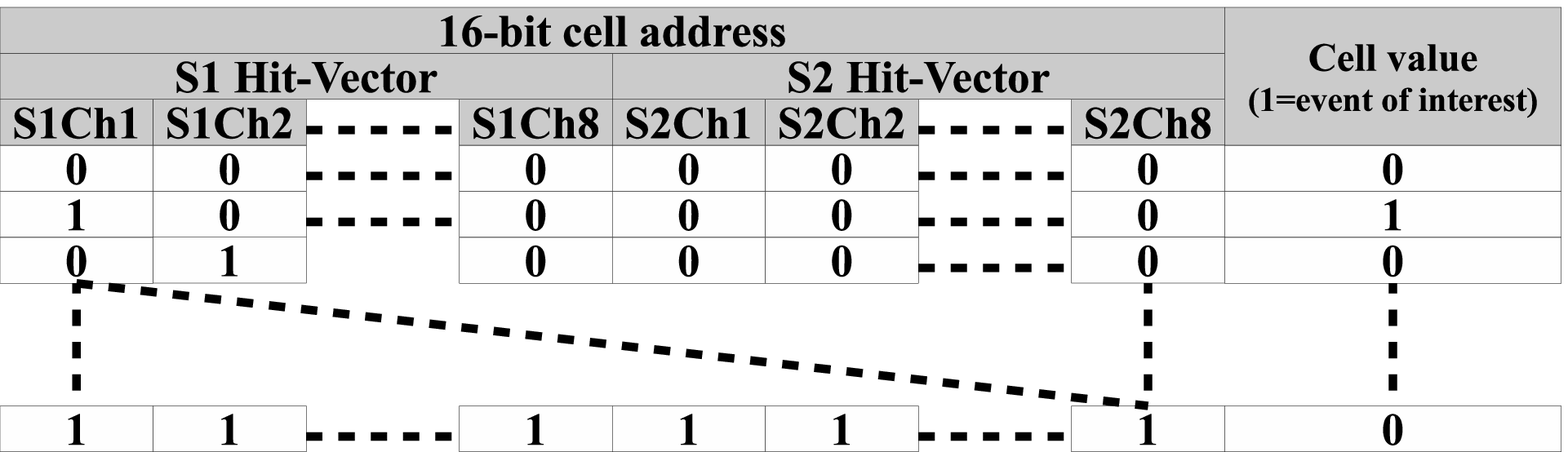}
\caption[TriggerMap]{Trigger Map at the DDC-8DSP level. The hit-vectors concatenated form a 16-bit address of the look-up table.}
\label{fig:TriggerMap}
\end{figure}

There are two levels of Trigger Maps (TM). One is at the DDC-8DSP and one is at the Trigger Builder level.
The Trigger Map on the DDC-8DSPs allows us to validate events based on the geometrical information provided by the S1 and S2 hit-vectors. Once the DDC-8DSP is done capturing the hit-vectors the S1 and S2 8-bit counterparts are concatenated to form a 16-bit address which is passed into the TM. It has $2^{16}$ single bit positions referenced by the just formed 16-bit address. The TMs bit positions are populated before the run by the end-user in a manner shown in Fig.~\ref{fig:TriggerMap}. A bit value of one indicates that the event with such a hit-address is of interest.\\\indent

\begin{figure}[!ht]
\centering
		\includegraphics[width=\figwidth\textwidth]{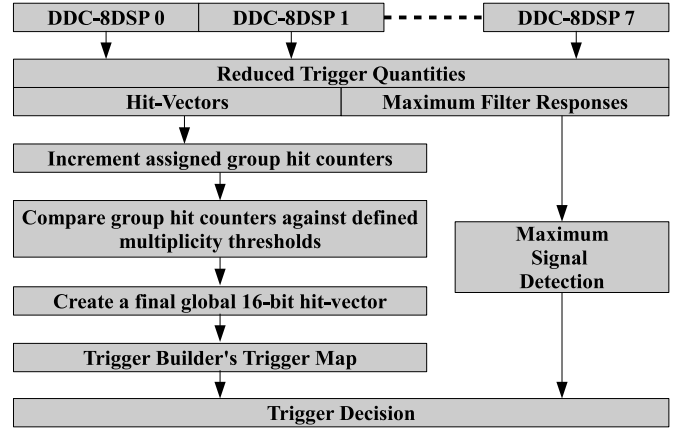}
\caption[TriggerFlow]{Trigger processing flow from the point where the DDC-8DSP modules send the reduced quantities, such as hit-vectors, maximum filter response, timestamps, to the TB where the final trigger decision is made.}
\label{fig:TriggerFlow}
\end{figure}

The TB was designed to work with up to seven DDC-8DSP modules. It can receive up to 56 bits worth of S1 or S2 hit-vectors. As a trigger map covering the potential $2^{56}$ combinations is beyond our memory resources, we developed a translation scheme, shown in Fig.~\ref{fig:TriggerFlow}. The up to 56-bit long hit-vectors are mapped into a secondary 16-bit hit-vector. Each of the received hit-bits is assigned to a hit-counter which represents the geometrical area to which a given group of PMTs was assigned. We scan through the hit-vectors and each time we see a logical one in a given cell, we find the hit counter tied to the given bit and increment it. This generates a sixteen cell count-vector that contain number of hits each area received. Based on user-defined counter thresholds we translate the count-vector into a final 16-bit hit-vector, which is passed to the TBs Trigger Map. It is worth noting that this translation module can be viewed as a very flexible multiplicity discriminator. 

\subsection{PMT summing and maximum-based trigger}
\label{subsec:PMTSummAndMaxTrigger}
\begin{figure}[!ht]
\centering
		\includegraphics[width=\figwidth\textwidth]{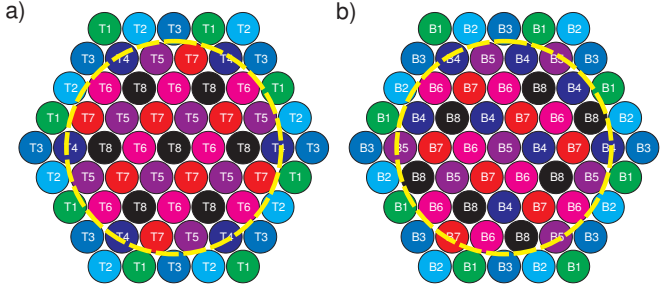}
\caption[PMTSumming]{Summing map for top (a) and bottom (b) PMTs. While scrambled to maximize the dynamic range, the grouping preserves the outer and inner PMT separation for event fiducialization. The fiducial volume boundary is indicated with the dashed-line circle.}
\label{fig:PMTSumming}
\end{figure}

\begin{figure}[!ht]
\centering
		\includegraphics[width=\figwidth\textwidth]{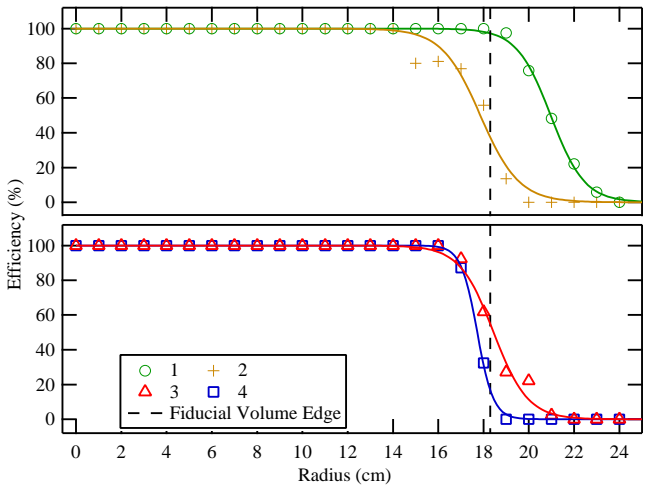}
\caption[MaximumTriggerEff]{Efficiency of accepting fiducial volume events: 1 - using threshold crossing optimized for large events (10,000 phe), 2 - using threshold crossing optimized for small events (7,500 phe), 3 - using maximum detection with T4 PMTs included, 4 - using maximum detection with T4 PMTs excluded. The solid lines are shown as eye-guides.}
\label{fig:MaximumTriggerEff}
\end{figure}

After analyzing many different summing approaches, the PMTs have been summed according to the map shown in Fig.~\ref{fig:PMTSumming}. For each PMT array (top and bottom) three sums (T1-T3, B1-B3) are dedicated to the outer PMTs and five (T4-T8, B4-B8) cover the PMTs above/below the fiducial volume, indicated in the figure with a dashed circle. PMTs are assigned to groups in such a way that no adjacent PMTs are in the same group, thus minimizing the saturation of the analog sum for large S2s. 
Using LUXSim~\cite{LUXSimPaper} we investigated the efficiency of event fiducialization capability of the trigger. We generated a set of events with known positions along the radius of the detector and applied different trigger algorithms. We found that using information only from threshold crossings has limitations, because this approach has to be optimized for ranges of energies. If one optimizes it for high energy events, smaller events from the outer volume will leak in, shown by curve 1 in Fig~\ref{fig:MaximumTriggerEff}. If one optimizes the thresholds for low energy events, then bigger fiducial volume events close to the boundary will be rejected, shown by curve 2 in Fig.~\ref{fig:MaximumTriggerEff}. 
To improve the triggering efficiency at the edge of the fiducial volume and minimize the dependency on event energy, we developed a maximum-based trigger on the FPGA. In this mode we keep track of which  trigger group sees the maximum filter response in a given event. The TM performs multiplicity discrimination and its output is combined with the maximum filter response to make a final trigger decision. For example if the Trigger Map determines that the coincidence condition is satisfied and the maximum signal was detected in one of the groups associated with the top PMTs above the fiducial volume, we accept a given event as valid, otherwise we reject it. 

\indent The maximum-based trigger improves the fiducialization efficiency, defined as deviation from the ideal cut at the FV edge, as shown by curve 3 in Fig.~\ref{fig:MaximumTriggerEff}. Because the FV edge cuts through the space covered by the inner-corner PMTs, we found that we can further improve the efficiency by summing them into one group in the top array (T4) and not taking this group into account when making the maximum detection based trigger, shown by curve 4 in Fig.~\ref{fig:MaximumTriggerEff}.

\section{Firmware and Software}
\begin{figure}[!ht]
\centering
		\includegraphics[width=\figwidth\textwidth]{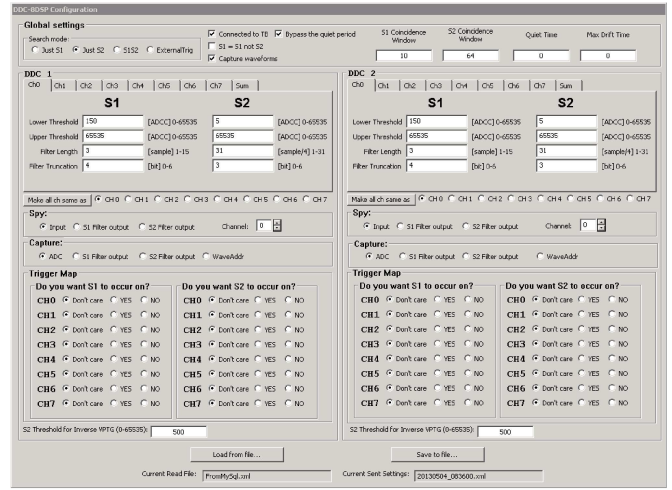}
\caption[Configuration]{Configuration menu of the DDC-8DSP modules. There are 20+ independent and configurable trigger parameters on the DDC-8DSP alone.}
\label{fig:Configuration}
\end{figure}

The FPGA firmware has been developed in Xilinx ISE environment~\cite{XilinxISE} using VHDL. The DDC-8DSP FPGA design consists of \midtilde5,000 lines of code and utilizes 27\% of the available resources. The Trigger Builder FPGA design consists of nearly 3,000 lines of code and utilizes 21\% of the available resources.
\\\indent
The FX USB controller firmware has been developed in $\mu$Vision3 ~\cite{uVision3} using C for 8051 processors. As the starting point we took a sample framework provided by Cypress Semiconductor~\cite{CypressFX2LP}. For the FX controller on DDC-8DSP we expanded it by adding just over 1,000 lines of code which almost fully utilizes the available resources. For the FX controller on the Trigger Builder we had to add just over 300 lines of code.
\\\indent
The USB communication has been realized using LibUSB~\cite{LibUsb}. It is an open-source USB communication library that allows cross-platform access to USB devices and has extended functionality such as read/write timeouts. In order to use the LibUsb library in the PC control software, a wrapper Dll library has been written in Dev-C++~\cite{DevCpp}.
\\\indent
The host PC control software is written in \textit{BlackBox Component Builder}. It is a Component Pascal development environment offered by Oberon Microsystems~\cite{BlackBoxOberon}. We have found this environment to be stable and rather friendly in rapid-prototyping. The implementation of the functionality for LUX took just above 3,000 lines of code. The developed GUI consist of two main and five support panels. Figure~\ref{fig:Configuration} shows one of the user configuration menus.
\\\indent
The triggering system is tightly coupled with the rest of the experiment via central MySQL databases. Whenever an DAQ acquisition is initiated, the trigger is re-programmed with the appropriate preset or custom settings which are saved in the database as an XML string~\cite{XMLSpec}. This aids record keeping and improves the integrity of collected data.
\\\indent 
The system constantly calculates an average trigger rate over ten second periods and reports these averages to the slow control system for storage and monitoring. This aids keeping track of the detector stability in time and allows operators to set additional alarm conditions.

\section{Performance}
\subsection{False Trigger Rates}
\begin{figure}[!ht]
\centering
		\includegraphics[width=\figwidth\textwidth]{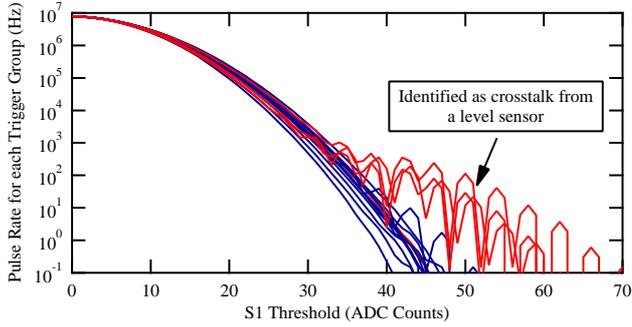}
\caption[TriggerSweepLevelSensor]{Trigger channel sweep showing crosstalk due to level sensors.}
\label{fig:TriggerSweepLevelSensor}
\end{figure}

It is critical to diagnose problems that impact the data being collected as soon as possible. To constantly monitor the trigger behavior we developed \textit{trigger sweeping}. It measures and records the rate of pulses seen by the S1 and S2 filters as a function of thresholds. It has been parallelized in such a way that it can run constantly in the background while the trigger system is operating and does not affect its performance. Figure~\ref{fig:TriggerSweepLevelSensor} shows one of the \textit{trigger sweeps} for the S1 filter, collected during one of our noise tests, that brought our attention to unexpected crosstalk from a liquid xenon level sensor.

\subsection{Trigger decision latency}
\label{sec:RespToSynthEvents}
The latency of the S1 Found signal, labeled by \textit{a} in Fig.~\ref{fig:SampleScopeEvent}, is associated with \textit{S1 = S1 not S2} being enabled and depends on the S1 and S2 filter lengths. The latency of 3~$\mu$s between an S2 being detected and the final trigger pulse is the sum of time of the programmed coincidence window in this case 2~$\mu$s, the time required to send reduced quantities to the TB, and the time to compute the final trigger decision, is shown by \textit{b} in Fig.~\ref{fig:SampleScopeEvent}.

%\begin{figure}[!ht]
%\centering
		%\includegraphics[width=0.48\textwidth]{SampleEventScopeTriggerV3.eps}
%\caption[SampleScopeEvent]{Generated S1\&S2 event and the trigger response. Ch4 - event waveform. Ch3 - signal indicating an S1 was found, in this case \textit{S1 = S1 not S2}, hence the detection delay. Ch2 - signal indicating an S2 was found. Ch1 - final trigger decision at the Trigger Builder level.}
%\label{fig:SampleScopeEvent}
%\end{figure}

\subsection{First WIMP Search}
\label{subsec:Run03WIMPSearch}
During the first WIMP search the trigger operated in \textit{S2Mode}.
In that mode, the S2 filter threshold was set to 8~phe on each of the trigger group channels and the coincidence was set to $\geq 2$ within a 2~$\mu$s time window. The hold-off was set to 1--4~ms, intentionally preventing additional triggers after large S2 pulses. 
Utilizing the trigger in this way minimized the chances of discarding potentially good events, while still offering significant savings in computational resources needed during event building.
The measured trigger efficiency reaches 99.9\% for S2 pulses with a total size of 100~phe~\cite{MongkolThesis}, which is a comfortable margin away from the 200~phe lower S2 cut in the experiments' first WIMP search analysis~\cite{LUXFirstResults}.

\subsubsection{Sample triggered low energy event}
\begin{figure}[!ht]
\centering
		\includegraphics[width=0.48\textwidth]{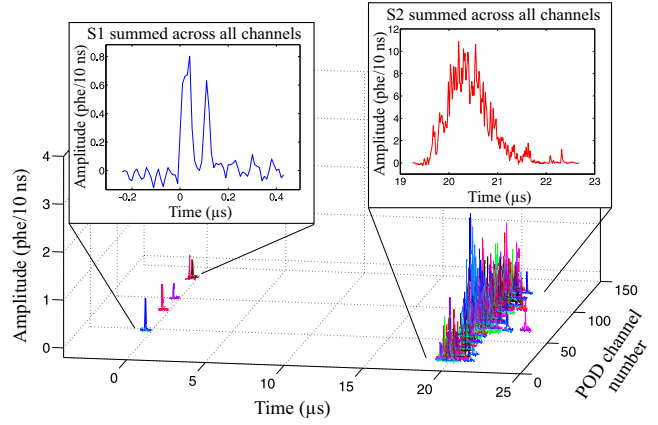}
\caption[Event1p5keVee]{A very small 1.5~keVee sample event in the LUX detector identified by the LUX trigger system.}
\label{fig:Event1p5keVee}
\end{figure}

The ability to capture very small energy-wise events is critical for success of LUX. Figure~\ref{fig:Event1p5keVee} shows a sample 1.5 keVee interaction event identified in the detector by the LUX trigger system during the first WIMP search.

\begin{figure}[!ht]
\centering
		\includegraphics[width=0.48\textwidth]{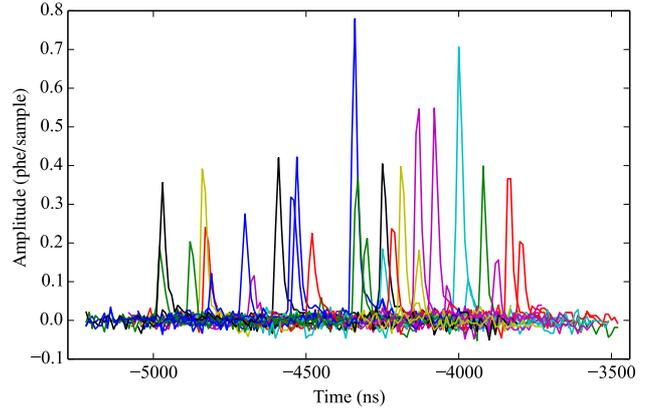}
\caption[SingleLeEvent]{An example of a single extracted electron with a total pulse area of only 24.6~phe which was detected by the LUX trigger system.}
\label{fig:SingleLeEvent}
\end{figure}

The smallest S2 signal in LUX is one induced by a single extracted electron from the liquid surface. Such a signal on average has a total size of 24.6~phe~\cite{LUXFirstResults}. Figure~\ref{fig:SingleLeEvent} shows that the LUX trigger system is capable of triggering on signals from single extracted electrons.

\subsection{Power consumption}

During off-line verification of the trigger system performance, where a software model of the trigger was developed~\cite{MongkolThesis}, it has been estimated that a software trigger would require a dedicated \midtilde150 CPU computational cluster to keep up with the deployed LUX trigger system. Such a dedicated cluster would require \midtilde6.7~kW to just power the CPUs themselves, while the LUX trigger boards consume a total of \midtilde15~W \cite{ErykThesis}.  

\section{Conclusions}

The LUX trigger is a powerful system that has been finely tuned and optimized for the LUX dark matter search experiment. 
This was achieved mainly by developing custom digital processing hardware and low-level FPGA firmware. Despite being forced to sum the PMT signals into sixteen trigger groups, the dynamic range has been optimized to accommodate low-energy dark matter searches, as well as high-energy detector calibrations. The trigger system has been shown to be sensitive to S2 signals induced by a single extracted electron with a latency of just few microseconds. Although the LUX experiment did not need to fully utilize the feature set of the developed trigger system, its flexibility and performance have shown to be invaluable at all the stages of the experiment. The system has been reliably operating since its underground deployment in early 2013 and currently continues to enable data collection during LUX experiments' 300-day run. The experience gained through this work has already shown to be important in the development of next-generation LZ experiment~\cite{LZCDR}.

\section*{Acknowledgments}

This work was partially supported by the U.S. Department of Energy (DOE) under award numbers DE-FG02-08ER41549, DE-FG02-91ER40688, DE-FG02-95ER40917, DE-FG02-91ER40674, DE-NA0000979, DE-FG02-11ER41738, DE-SC0006605, DE-AC02-05CH11231, DE-AC52-07NA27344, and DE-FG01-91ER40618; the U.S. National Science Foundation under award numbers PHYS-0750671, PHY-0801536, PHY-1004661, PHY-1102470, PHY-1003660, PHY-1312561, PHY-1347449; the Research Corporation grant RA0350; the Center for Ultra-low Background Experiments in the Dakotas (CUBED); and the South Dakota School of Mines and Technology (SDSMT). LIP-Coimbra acknowledges funding from Funda\c{c}\~{a}o para a Ci\^{e}ncia e a Tecnologia (FCT) through the project-grant CERN/FP/123610/2011. Imperial College and Brown University thank the UK Royal Society for travel funds under the International Exchange Scheme (IE120804). The UK groups acknowledge institutional support from Imperial College London, University College London and Edinburgh University, and from the Science \& Technology Facilities Council for PhD studentship ST/K502042/1 (AB). The University of Edinburgh is a charitable body, registered in Scotland, with registration number SC005336. 

We gratefully acknowledge the logistical and technical support and the access to laboratory infrastructure provided to us by the Sanford Underground Research Facility (SURF) and its personnel at Lead, South Dakota. SURF was developed by the South Dakota Science and Technology authority, with an important philanthropic donation from T. Denny Sanford, and is operated by Lawrence Berkeley National Laboratory for the Department of Energy, Office of High Energy Physics.

%\subsection{Single photoelectron spectrum}
%Figure~\ref{fig:SinglePhePMT20} shows an example of a measured single photoelectron spectrum as an average of all the channels seen by the trigger modules.
%This real measurement shows that since we are summing several PMT signals into one trigger channel, we can capture 85.2\% of single photoelectrons at a threshold that induces ~1 Hz of False Triggers. This measured result is in great agreement with what was predicted in the earlier discussed simulations (Section~\ref{sec:Filtersensitivity}).
%\begin{figure}[!ht]
%\centering
		%\includegraphics[width=\figwidth\textwidth]{SinglePheSpectrum.eps}
%\caption[SinglePhePMT20]{Single-phe spectrum as seen by the DDC-8DSP module}
%\label{fig:SinglePhePMT20}
%\end{figure}

%%\clearpage

%\section*{References}

\end{document}